\documentclass[prd,superscriptaddress,twocolumn,nofootinbib,showpacs,preprintnumbers,floatfix]{revtex4}

\usepackage{mathrsfs}
\usepackage{amsfonts,amssymb,amsmath}
\usepackage{graphicx,epsfig}
\usepackage{multirow}
\usepackage{verbatim}

\def\fsu5{$\cal{F}$-$SU(5)$}
\def\bfsu5{$\boldsymbol{\mathcal{F}}$-$\boldsymbol{SU(5)}$}

\def\m1half{$M_{1/2}$}
\def\m3half{$M_{3/2}$}
\def\m32{$M_{32}$}

\def\fb{${\rm fb}^{-1}$~}

\def\mt2{$M_{T2}$}
\def\x2{$\chi^2$}
\def\2b{$M_{T2}b$}
\def\sb{$S/\sqrt{B+1}$~}
\def\bs0{$B_S^0 \rightarrow \mu^+ \mu^-$}

\begin{document}

\title{A 125.5 GeV Higgs Boson in \bfsu5: Imminently Observable Proton Decay, A\\130 GeV Gamma-ray Line, and SUSY Multijets \& Light Stops at the LHC8}

\author{Tianjun Li}

\affiliation{State Key Laboratory of Theoretical Physics and Kavli Institute for Theoretical Physics China (KITPC),
Institute of Theoretical Physics, Chinese Academy of Sciences, Beijing 100190, P. R. China}

\affiliation{George P. and Cynthia W. Mitchell Institute for Fundamental Physics and Astronomy, Texas A$\&$M University, College Station, TX 77843, USA}

\author{James A. Maxin}

\affiliation{George P. and Cynthia W. Mitchell Institute for Fundamental Physics and Astronomy, Texas A$\&$M University, College Station, TX 77843, USA}

\author{Dimitri V. Nanopoulos}

\affiliation{George P. and Cynthia W. Mitchell Institute for Fundamental Physics and Astronomy, Texas A$\&$M University, College Station, TX 77843, USA}

\affiliation{Astroparticle Physics Group, Houston Advanced Research Center (HARC), Mitchell Campus, Woodlands, TX 77381, USA}

\affiliation{Academy of Athens, Division of Natural Sciences, 28 Panepistimiou Avenue, Athens 10679, Greece}

\author{Joel W. Walker}

\affiliation{Department of Physics, Sam Houston State University, Huntsville, TX 77341, USA}


\begin{abstract}

We establish that the light Higgs boson mass in the context of the No-Scale Flipped $SU(5)$ GUT with TeV
scale vector-like matter multiplets (flippons) is consistent with $m_h = 125.5\pm0.5$ GeV in the region of the best
supersymmetry (SUSY) spectrum fit to low statistics data excesses observed by ATLAS in multijet and
light stop 5 \fb SUSY searches at the LHC7.  Simultaneous satisfaction of these disparate goals is
achieved by employing a minor decrease in the $SU(5)$ partial unification scale $M_{32}$ to lower
the flippon mass, inducing a larger Higgs boson mass shift from the flippon loops.
The reduction in $M_{32}$, which is facilitated by a phenomenologically favorable reduction of the
low-energy strong coupling constant, moreover suggests an imminently observable ${(e\vert\mu)}^{\!+}\! \pi^0$
proton decay with a central value time scale of $1.7\times10^{34}$~years.  At the same point in the model space, we find a lightest neutralino
mass of $m_{\chi} = 145$ GeV, which is suitable for the production of 130 GeV monochromatic gamma-rays
through annihilations yielding associated $Z$-bosons; a signal with this energy signature has been
identified within observations of the galactic center by the FERMI-LAT Space Telescope.
In conjunction with direct correlations to the fate of the ATLAS multijet and light stop production channels
presently being tested at the LHC8, we suggest that the reality of a 125.5~GeV Higgs boson affords 
a particularly rich company of specific and imminently testable associated observables. 

\end{abstract}



\preprint{ACT-11-12, MIFPA-12-29}

\maketitle


\section{Introduction}

The recent combined discovery of the CP-even light Higgs boson around $m_h = 125$-$126$~GeV by the
ATLAS~\cite{ATLAS:2012gk}, CMS~\cite{CMS:2012gu}, and CDF/D0~\cite{Aaltonen:2012qt}
Collaborations has sparked detailed examinations into what regions of the SUSY parameter space remain
viable, and which are very constrained, if not outright excluded.  Motivated by these new precision
experimental measurements, we endeavor here to commensurately enhance the precision of prior theoretical
estimates~\cite{Li:2011xg,Li:2011ab} for the range of Higgs boson masses consistent with
the No-Scale Flipped $SU(5)$ model with vector-like matter (flippons)~\cite{Li:2010ws,Li:2010mi,Li:2010uu,
Li:2011dw,Maxin:2011hy,Li:2011hr,Li:2011xu,Li:2011in,Li:2011gh,Li:2011rp,
Li:2011fu,Li:2011xg,Li:2011ex,Li:2011av,Li:2011ab,Li:2012hm,Li:2012tr,Li:2012ix,Li:2012yd,Li:2012qv},
dubbed \fsu5 for short.  The mutual consistency of a $125$-$126$~GeV Higgs boson mass with
an explanation for the tantalizing positive excesses~\cite{Li:2012tr,Li:2012ix}
observed at low statistics by ATLAS~\cite{ATLAS-CONF-2012-033,ATLAS-CONF-2012-037} and CMS~\cite{SUS-12-002} in
the $\sqrt{s} = 7$ TeV SUSY search dramatically narrows the \fsu5 model space, with interesting
and specific implications for ongoing collider, indirect dark matter detection, and proton decay experiments. 

The central mass peak of the new particle observed at the LHC exists at 125.3 GeV and 126.0 GeV,
according respectively to the CMS~\cite{CMS:2012gu} and ATLAS~\cite{ATLAS:2012gk} collaborations.
In each case, statistical and systematic errors of about half a GeV are suggested.
Lacking an official combination of the two experiments' statistics, a more broad treatment might follow something akin to
the proposal $m_h = 125.0 \pm 1.0~{\rm (exp)} \pm 1.5~{\rm (theory)}$~GeV of Ref.~\cite{Buchmueller:2011ab}, noting
in particular that our capacity to computationally model the predicted Higgs value is likewise imprecise. 
However, it is our interest in the present work to ascertain the restrictions that a Higgs mass very close to
the average of centrally reported values would place on an otherwise successful model.  Insomuch as this mass is
somewhat heavier than what may be comfortably achieved in typical GUT models without resorting to very heavy
sparticles, we are interested in demonstrating a counter-example satisfying that vital emerging constraint
which does not resort to taking extremities in the admissible lower bound.  In particular, we presently
investigate the possibility that a large portion of the burden could be absorbed (in conjunction with the described
flippon loops) by a lowering of the strong coupling $\alpha_{\rm s}$ at $M_{\rm Z}$, an accommodation to which the
flipped $SU(5)$ GUT is particularly well historically adapted~\cite{Ellis:1995at}.  The Higgs mass constraint
$m_h = 125.5\pm0.5$ adopted for this work is thus purposefully rather strict.

\section{\bfsu5: The Model}

No-Scale \fsu5 is built upon the foundation of the Flipped
$SU$(5)~\cite{Barr:1981qv,Derendinger:1983aj,Antoniadis:1987dx} Grand Unified Theory (GUT), two
pairs of hypothetical TeV-scale flippon multiplets of mass $M_V$ derived from local F-Theory model
building~\cite{Jiang:2006hf,Jiang:2009zza,Jiang:2009za,Li:2010dp,Li:2010rz}, and the
dynamically established boundary conditions of No-Scale
supergravity~\cite{Cremmer:1983bf,Ellis:1983sf, Ellis:1983ei, Ellis:1984bm, Lahanas:1986uc}.
In flipped $SU(5)\times U(1)_X$ models, there are two unification scales: the $SU(3)_C\times SU(2)_L$
unification scale $M_{32}$ and the $SU(5)\times U(1)_X$ unification scale $M_{\cal F}$. To separate the
$M_{32}$ and $M_{\cal F}$ scales and obtain true string-scale gauge coupling unification in free
fermionic string models~\cite{Jiang:2006hf, Lopez:1992kg} or the decoupling scenario in F-theory
models~\cite{Jiang:2009zza, Jiang:2009za}, we introduce vector-like particles, called flippons,
which form complete flipped $SU(5)\times U(1)_X$ multiplets. In the most elementary No-Scale scenario,
$M_0 = A = B_{\mu} = 0$ at the ultimate unification scale $M_{\cal F}$, while the entire group of low energy
SUSY breaking soft-terms evolve down from just one non-zero parameter $M_{1/2}$. As a result, the
particle spectrum is proportional to $M_{1/2}$ at leading order, rendering the bulk ``internal''
physical properties invariant under an overall rescaling. The matching condition between the
low-energy value of $B_\mu$ that is demanded by electroweak symmetry breaking (EWSB) and the high-energy
$B_\mu = 0$ boundary condition is quite difficult to reconcile under the renormalization group equation (RGE)
running. The solution at hand applies modifications to the $\beta$-function coefficients that are
produced by the flippon loops. Naturalness in view of the gauge hierarchy and $\mu$ problems suggest that
the flippon mass $M_{V}$ should be of the TeV order. Avoiding a Landau pole for the strong coupling constant
restricts the set of vector-like flippon multiplets which may be given a mass in this range to only two
constructions with flipped charge assignments, which have been explicitly realized in the $F$-theory
model building context~\cite{Jiang:2006hf,Jiang:2009zza, Jiang:2009za}. In either case, the
(formerly negative) one-loop $\beta$-function
coefficient of the strong coupling $\alpha_3$ becomes precisely zero from $M_{32}$ to $M_{V}$,
flattening the RGE running of the gaugino mass $M_3$ and the strong coupling $\alpha_3$
between these two scales. The wide residual gap which this leaves between the $\alpha_{5}$ and $\alpha_{\rm X}$
couplings at $M_{32}$ facilitates a quite important secondary running phase up to the final $SU(5) \times U(1)_{\rm X}$
unification scale, which may be elevated by 2-3 orders of magnitude into adjacency with the Planck mass, where the $B_\mu = 0$
boundary condition fits like hand to glove~\cite{Ellis:2001kg,Ellis:2010jb,Li:2010ws}.

\section{Higgs Boson Mass Shift}

The correlation of the predicted \fsu5 Higgs boson mass in Ref.~\cite{Li:2011ab} with the collider
measured value was achieved via contributions to the lightest Higgs boson from the vector-like flippons,
computed from the RGE improved one-loop effective Higgs potential
approach~\cite{Babu:2008ge,Martin:2009bg}, resulting in a 3-4 GeV upward shift in the Higgs boson mass
to the experimentally measured range. The relevant shift possesses a leading dependence on the flippon
mass $M_V$, with larger shifts corresponding to lighter vector-like flippons. The recently improved
measurements that further constrain the Higgs boson mass to within a more narrow corridor of $m_h = 125.5
\pm 0.5$ GeV necessitate a refinement in our calculations from our previous \fsu5 estimate of $m_h \simeq
124$ GeV~\cite{Li:2012tr,Li:2012qv} at that location in the model space capable of explaining the
multijet and light stop excesses. In order to accomplish such, the flippon mass $M_V$ must be lowered
further to accommodate a larger flippon contribution to the Higgs boson mass.

A multi-axis \x2 test was applied to those ATLAS and CMS searches exhibiting the largest production of
events beyond the data-driven Standard Model predictions within the high-energy framework of
\fsu5~\cite{Li:2012tr,Li:2012ix}. It was shown that the best fit SUSY mass to these searches occurred at
$M_{1/2} \simeq 700$ GeV. The No-Scale \fsu5 SUSY spectrum possesses the rather distinct characteristic
of leading order $en~ masse$ proportionality to only the single dimensionful parameter $M_{1/2}$. In
essence, the internal physics of \fsu5 are largely invariant under a numerical rescaling of only
$M_{1/2}$. For this reason, we must leave $M_{1/2}$, which is defined at the ultimate $SU(5) \times
U(1)_X$ unification scale $M_{\cal F}$, relatively unchanged, while simultaneously shifting the
flippon mass $M_V$, since the event landscape at the LHC is chiefly correlated to $M_{1/2}$. This
rescaling of the mass scale $M_{1/2}$ shares an analogous historical bond with the fixing of the Bohr
atomic radius $a_0 = 1/(m_e \alpha)$ in terms of the physical electron mass and charge, through
minimization of the electron energy~\cite{Feynman}. In each of these two instances, the spectrum scales
according to variation in the selected constants, while leaving the relative internal structure of the
model intact.

The key to realizing a modest shift in $M_V$ while maintaining stability in $M_{1/2}$ at the ultimate
unification scale $M_{\cal F}$ is the recognition that some flexibility remains with regards to the
positioning of the penultimate $SU(3)_C \times SU(2)_L$ unification scale $M_{32}$. By
subjecting $M_{32}$ to minor variations, we can deviate the flippon mass $M_V$ through the RGE running,
while preserving reasonable stability of $M_{1/2}$ at $M_{\cal F}$.  In practice, small transitions in
the $M_{32}$ scale are executed in reverse by programmatic variation of the low-energy strong coupling
constant $\alpha_{\rm s}$. The strong coupling constant fulfills a low-energy boundary condition for the RGE
running, thus small movements at low-energy, while keeping $M_{1/2}$ at $M_{\cal F}$ reasonably fixed,
translate into minor shifts in $M_{32}$. These small variations in the low and high-energy boundary
conditions thus map onto the desired motion of the flippon mass $M_V$, while maintaining the
favorable phenomenology dictated by a particular $M_{1/2}$.

To quantify the allowable latitude of variation for the low-energy strong coupling constant $\alpha_{\rm s}$, we
turn to the most recent report of precision electroweak scale measurements of this parameter~\cite{Bandurin:2011sh},
which announces the refined result $\alpha_{\rm s} = 0.1161^{+0.0041}_{-0.0048}$. Considering that our previous work in its
entirety~\cite{Li:2010ws,Li:2010mi,Li:2010uu,Li:2011dw,Maxin:2011hy,Li:2011hr,Li:2011xu,Li:2011in,Li:2011gh,Li:2011rp,
Li:2011fu,Li:2011xg,Li:2011ex,Li:2011av,Li:2011ab,Li:2012hm,Li:2012tr,Li:2012ix,Li:2012yd,Li:2012qv}
employed a value of $\alpha_{\rm s} = 0.1172$, quite soundly within the measured QCD uncertainties, though a bit
above the centrally measured value, we may presently take advantage of an opportunity to displace this coupling
downward, toward and through the central value.  In fact, shifts in $\alpha_{\rm s}$ and $M_{32}$
operate with a parallel directionality, inducing a flow of the flippon mass $M_V$ that is likewise of the
same sign.  Hence, the small shortfall between our previous Higgs boson mass calculation of $m_h \simeq 124$ GeV
and the now more highly constrained Higgs boson mass measurements centered at $125.5$~GeV may be attributed to the
application of an overly large strong coupling constant $\alpha_{\rm s}$ at the low-energy boundary condition.

As we shall explicitly show, lowering $\alpha_{\rm s}$ to and slightly below the recently reported central value does
indeed permit a sufficient decrease in the flippon mass $M_V$ to allow an $m_h \simeq 125.5$ GeV Higgs boson
mass at the well of our cumulative $\chi^2$ best fit to the multijet and light stop excesses at the LHC.
Moreover, oscillating the strong coupling constant $\alpha_{\rm s}$ around its central empirically
derived value in turn oscillates the \fsu5 computed Higgs boson mass around $m_h \sim 125$ GeV,
prominently emphasizing the natural and non-trivial \fsu5 SUSY GUT correlation between two independent high-precision
low-energy experimental measurements, specifically the strong coupling constant $\alpha_{\rm s}$ and light
Higgs boson mass $m_h$.  In Section~(\ref{sct:protons}), we will supplement numerically based observations on
the co-dependency of $M_{32}$ and $\alpha_{\rm s}$ with an analytical treatment emphasizing the vital
role that lowering the GUT scale $M_{32}$ plays in hastening another key indicator of post Standard Model physics:
the decay of the proton.

\section{Higgs Boson Mass Calculations}

We vary the strong coupling constant $\alpha_{\rm s}$ to ascertain those regions of the model space capable
of handily generating an $m_h = 125.5 \pm 0.5$ GeV Higgs boson mass. To first constrain the model space to
only those regions satisfying current experiment, we apply the set of stable ``bare-minimal''
experimental constraints of Ref.~\cite{Li:2011xu}, consisting of the top quark mass $172.2~{\rm GeV}
\leq m_{\rm t} \leq 174.4~{\rm GeV}$, 7-year WMAP cold dark matter relic density $0.1088 \leq \Omega_{\rm
CDM}h^2 \leq 0.1158$~\cite{Komatsu:2010fb}, and precision LEP constraints on the SUSY mass content. In
the bulk of the surviving model space the lightness of the stau, which is itself a potential future target
for direct collider probes by the forthcoming $\sqrt{s} = 14$~TeV LHC, is leveraged to facilitate an
appropriate dark matter relic density via stau-neutralino coannihilation. The SUSY particle masses and
relic densities are calculated with {\tt MicrOMEGAs 2.4}~\cite{Belanger:2010gh}, via application of a
proprietary modification of the {\tt SuSpect 2.34}~\cite{Djouadi:2002ze} codebase to evolve the
flippon-enhanced RGEs. The resultant model space is then used as a basis to compute the flippon
contribution to the Higgs boson mass per Ref.~\cite{Martin:2009bg}, as a first approximation in
deriving the total Higgs boson mass $m_h$. This carves out a very narrow strip of model space as a function of
$(M_{1/2}, M_V)$ that will shift as a unit along the $M_V$ axis in response to small variations in $\alpha_{\rm s}$. These narrow strips
of model space are subsequently utilized to complete more detailed calculations of the Higgs boson mass to
be described shortly. We find an $M_{32} \simeq 7 \times 10^{15}$ GeV and $\alpha_{\rm s} = 0.1145$ to be ideal
for achieving a Higgs boson mass of $m_h \sim 125.5$ GeV.

A brief comparison is in order between the structure of the vector-like matter
content appearing in the models of Ref.~\cite{Martin:2009bg} and that employed in \fsu5.
The \fsu5 construction adopts the multiplets
\begin{eqnarray}
& {XF} = {\mathbf{(10,1)}} \equiv (XQ,XD^c,XN^c) ~;~ {\overline{XF}} = {\mathbf{({\overline{10}},-1)}} & \nonumber \\
& {Xl} = {\mathbf{(1, -5)}} \quad;\quad{\overline{Xl}} = {\mathbf{(1, 5)}}\equiv XE^c \, , &
\label{eq:flippons}
\end{eqnarray}
where $XQ$, $XD^c$, $XE^c$ and $XN^c$ carry the same quantum numbers as the quark doublet, right-handed down-type quark,
charged lepton and neutrino, respectively.
This is one of two possibilities realizing complete $SU(5)\times U(1)_X$ multiplets at the TeV
scale that avoids a Landau pole for the strong coupling constant; alternatively, 
the pair of $SU(5)$ singlets may be discarded, but phenomenological consistency then
requires the substantial application of unspecified GUT thresholds.  Both scenarios
have been explicitly realized in the $F$-theory and free-fermionic model building
contexts~\cite{Jiang:2006hf,Jiang:2009zza, Jiang:2009za}.
The $SU(3)_C \times SU(2)_L$ gauge symmetry unification is not consequently shifted far 
from the traditional GUT scale around $10^{16}$ GeV, although the vector-like multiplets
do elevate the $SU(5)\times U(1)_X$ gauge symmetry unification into close proximity with
the string scale.

In Ref.~\cite{Martin:2009bg}, the author has considered vector-like matter corresponding to
the charge assignments
\begin{equation}
(XQ, XQ^c) + (XD, XD^c) + 2 (XE, XE^c)\, ,
\end{equation}
which do not form complete $SU(5)$ multiplets,
but which nevertheless do preserve a traditional GUT coupling unification scale.
The detailed string-theoretic construction of GUTs with such vector-like particles
remains an open question, so this model is classified as phenomenological.  Nevertheless,
the Yukawa superpotential terms
\begin{equation}
XQ^c XD H_u \quad \& \quad XQ XD^c H_d
\end{equation}
employed in Ref.~\cite{Martin:2009bg} to increase the lightest CP-even Higgs boson
mass are directly applicable to the \fsu5 field content, such that the discussions
and formulae established in that work are likewise directly relevant to our own analysis.

The mechanism for generation of the flippon contributions to the Higgs boson mass is the
following pair of Yukawa interaction terms between the MSSM Higgs boson and the vector-like flippons in
the superpotential.
\begin{equation}
W = {\frac{1}{2}} Y_{xd} \, XF \, XF \, h + {\frac{1}{2}} Y_{xu} \, \overline{XF} \, \overline{XF} \,
\overline{h}
\end{equation}
Being vector-like rather than chiral, the flippons are also afforded a proper ``diagonal'' Dirac mass.
After the $SU(5)\times U(1)_X$ gauge symmetry is broken down to the Standard Model, the relevant Yukawa
couplings are
\begin{equation}
W = Y_{xd} XQ XD^c H_d + Y_{xu} XQ^c XD H_u~.
\end{equation}
We employ the RGE improved one-loop effective Higgs potential approach to calculate the contributions to
the lightest CP-even Higgs boson mass from the vector-like
particles~\cite{Babu:2008ge,Martin:2009bg}. The relevant shift in the Higgs boson mass-square is
approximated as~\cite{Huo:2011zt}
\begin{eqnarray}
\Delta m_h^2 &=& -\frac{N_c M_Z^2}{8\pi^2}\times \cos^22\beta~({\hat Y}_{xu}^2+{\hat Y}_{xd}^2)t_V
\nonumber \\
&+&\frac{N_cv^2}{4\pi^2}\times{\hat Y}_{xu}^4~(t_V+\frac{1}{2}X_{xu})~,
\label{Delta mhs}
\end{eqnarray}
with
\begin{eqnarray}
&{\hat Y}_{xu}=Y_{xu}\sin\beta \quad;\quad {\hat Y}_{xd}=Y_{xd}\cos\beta&
\nonumber\\
&{\tilde A}_{xu}=A_{xu}-\mu\cot\beta \quad;\quad t_V=\ln\frac{M_S^2+M_V^2}{M_V^2}&
\\
&X_{xu}=\frac{ -2M_S^2(5M_S^2+4M_V^2) + 4(3M_S^2 - 2M_V^2) {\tilde A}_{xu}^2+{\tilde A}_{xu}^4}{6(M_V^2+M_S^2)^2}&~,
\nonumber
\label{eq:vectorhiggs}
\end{eqnarray}
where $N_c$ is the number of colors carried by the vector-like fields, $M_S$ is the geometric mean of the
left-handed and right-handed stop masses at low energy, and ${A}_{xu}$ is the soft SUSY breaking
trilinear term for the Yukawa superpotential element $Y_{xu} XQ^c XD H_u$.

\begin{figure*}[htp]
        \centering
        \includegraphics[width=1.00\textwidth]{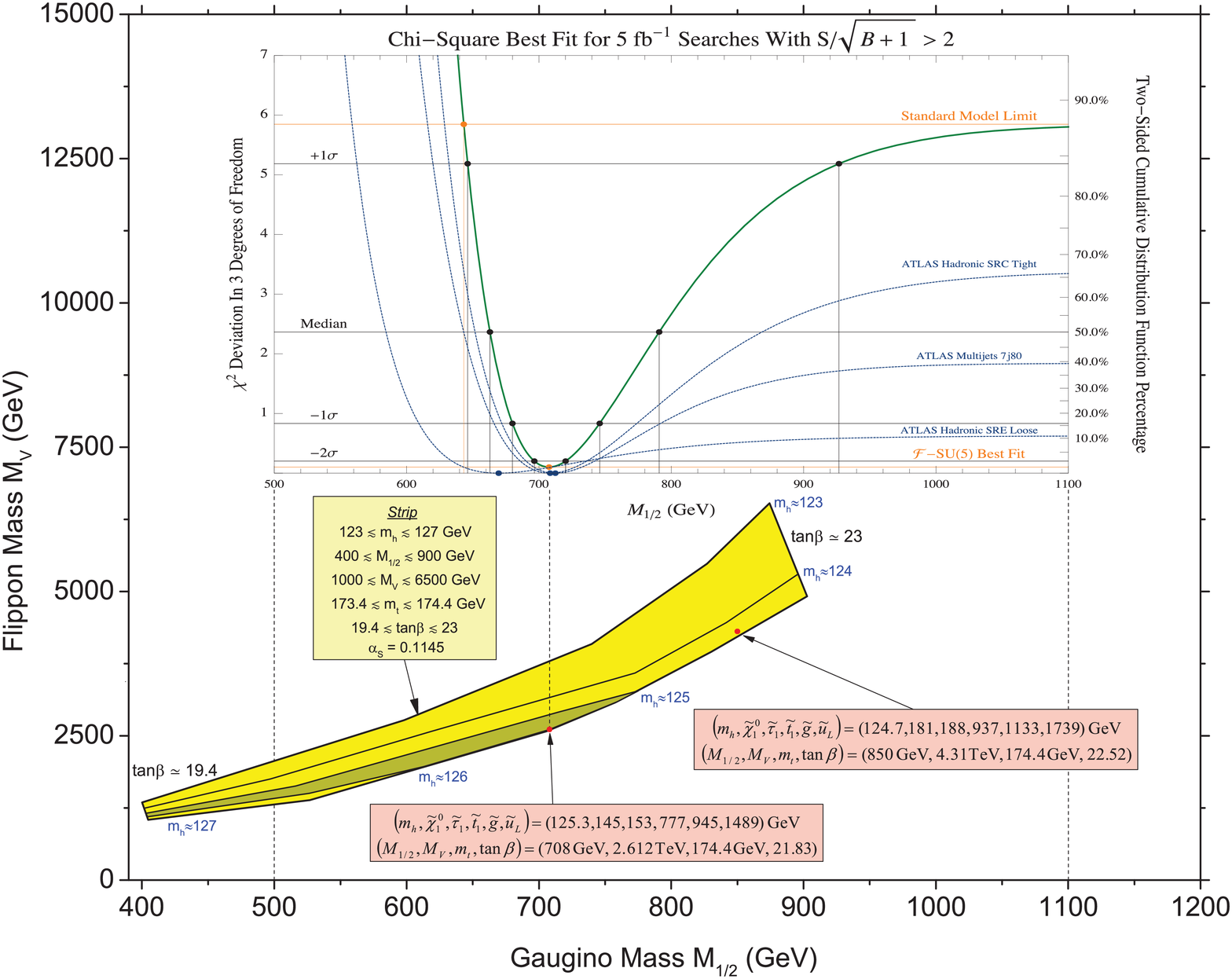}
        \caption{We depict the experimentally viable parameter space of No-Scale \fsu5 as a function of the gaugino mass $M_{1/2}$
and flippon mass $M_V$. The surviving model space after application of the bare-minimal constraints of
Ref.~\cite{Li:2011xu} and Higgs boson mass calculations is illustrated by the yellow strip, representing $123 \lesssim
m_h \lesssim 127$ GeV. A visual indication of the freedom that may be realized by relaxing the strong bounds on the Higgs mass is
given by inclusion of outer contours at 123 and 124 GeV, corresponding to a weaker interpretation of the Higgs
mass central value and computational uncertainties, and an outer contour at 127 GeV representing a stronger interpretation of the central value. The inset diagram (with linked horizontal scale) is the multi-axis cumulative \x2
fitting of Ref.~\cite{Li:2012tr}, though here using a slightly altered strong coupling constant of
$\alpha_{\rm s} = 0.1145$. Additionally, updates to certain of the relevant published~\cite{ATLAS-CONF-2012-033}
background estimates have actually enhanced the agreement in the mass scale favored across the displayed
search channels. Highlighted are benchmarks at $M_{1/2} = 708$ GeV and $M_{1/2} = 850$ GeV; the complete SUSY mass spectra are given in Tables~\ref{tab:masses} and~\ref{tab:massestwo}.}
        \label{fig:strip}
\end{figure*}

The Yukawa couplings $Y_{xu} $ and $Y_{xd} $ are 
required to be smaller than about 3 throughout the full running up 
to the unification scale due to the perturbative bound.
Thus, they are smaller than about 1.02 at low energies,
as detailed in Ref.~\cite{Huo:2011zt}.
Specifically, from the two-loop RGE analyses, we determined that the maximal Yukawa couplings $Y_{xu}$ are
about 0.96, 1.03, and 1.0 for $\tan\beta=2$, $\tan\beta \sim 23$, and $\tan\beta=50$,
respectively~\cite{Huo:2011zt}, and thus choose a working value of $Y_{xu}=1.0$.
Note, in particular, that the Yukawa couplings $Y_{xu}$ and $Y_{xd}$
remain fully distinct, even in the flipped $SU(5) \times U(1)_X$ models.
Thus, for simplicity, we here assume that $Y_{xd} =0$.
With both $Y_{xu}$ and $Y_{xd}$ non-zero, one may study
the RGE running and Higgs boson mass similarly, and the established 
results for the Higgs boson mass are found to broadly hold~\cite{Huo:2011zt}.
The corrected Higgs
boson mass will be
\begin{eqnarray}
m_h ~=~ \sqrt{({\widetilde{m}}_h)^2 +\Delta m_h^2}~,~\,
\label{eq:higgs}
\end{eqnarray}
where $\widetilde{m}_h$ is the Higgs boson mass in the Minimal Supersymmetric Standard Model (MSSM).
We implement this detailed Higgs boson mass computational methodology on points sampled from the residual strip of model
space surviving subsequent to application of the ``bare-minimal'' constraints~\cite{Li:2011xu} and the simplified first-level Higgs boson
mass approximation detailed in Ref.~\cite{Li:2011ab}.  In that preliminary analysis, a formula for the Higgs
mass-square shift $\Delta m_h^2$ attributable to the vector-like flippon multiplets (and also more weakly to the soft term
mass approximated by $M_S \simeq 2 M_{1/2}$) is adopted that implements the leading mass
scale dependencies identified in Section~III of Ref.~\cite{Martin:2009bg}.
This allows us to reasonably extrapolate a detailed computation of the corrected Higgs mass across
the \fsu5 model space, isolating a narrow region of
the model space supporting $123 \lesssim m_h \lesssim 127$ GeV, as illustrated in Figure (\ref{fig:strip}).
\begin{eqnarray}
{m}_h &\simeq& \sqrt{ \widetilde{m}_h^2 + (87.81~{\rm GeV})^2 \times \left( \ln x - \frac{5}{6} +\frac{1}{x} -\frac{1}{6x^2} \right)}
\nonumber \\
x &\equiv& 1 + {\left( \frac{2 M_{1/2}}{M_{\rm V}} \right)}^2
\label{eq:m_higgs}
\end{eqnarray}
Larger shifts are seen to correspond to lighter vector-like multiplets.
This effect works in tandem with the top quark mass, which varies inversely with $M_{\rm V}$ in the viable \fsu5 model space~\cite{Li:2011ab},
and whose elevation likewise lifts the bare Higgs mass prediction $\widetilde{m}_h$.
For both reasons, larger net values of the Higgs mass $m_h$ occur toward the lower \fsu5 model boundary for $M_{\rm V}$, just prior to the
extremity of a single standard deviation from the top quark world average.

Note that no 5 \fb 7 TeV LHC constraints are imposed upon the strip in Figure (\ref{fig:strip}). The
enduring region is comprised of a narrow strip of space confined to 400 $\lesssim M_{1/2} \lesssim$ 900
GeV, 19.4 $\lesssim$ tan$\beta$ $\lesssim$ 23, and 1000 $\lesssim M_V \lesssim$ 6500 GeV. The border at
the minimum $M_{1/2} \simeq$ 400 GeV is required by the LEP constraints, while the maximum boundary at
$M_{1/2} \simeq$ 900 GeV prevents a charged stau LSP. We superimpose upon the plot
space in Figure (\ref{fig:strip}) the cumulative \x2 curve for a set of ATLAS SUSY searches
exhibiting appreciable signal significance over the expected background in accordance with the procedures detailed in
Refs.~\cite{Li:2012tr,Li:2012ix}.  The reduction in the strong coupling constant to
$\alpha_{\rm s} = 0.1145$ is carried over into this analysis. Figure (\ref{fig:strip}) depicts the successful
preservation of a best SUSY mass fit at $M_{1/2} \simeq 700$ GeV, in tandem with a smaller $M_V \simeq 2600$ GeV,
which provides the desired upward shift in the Higgs boson mass to $m_h \simeq 125.5$ GeV.
Accordingly, the SUSY mass spectrum of the benchmark model of Table~\ref{tab:masses} at the \x2 well
remains rather similar to that of Ref.~\cite{Li:2012tr}, with a bino-dominated LSP
$m_{\widetilde{\chi}_1^0} = 145$~GeV, light stop $m_{\widetilde{t}_1} = 777$~GeV, gluino
$m_{\widetilde{g}} = 945$~GeV, and heavy squark $m_{\widetilde{u}_L} = 1489$ GeV. We also provide a somewhat heavier spectrum in Table~\ref{tab:massestwo}, as graphically annotated in Figure (\ref{fig:strip}), that is within one standard deviation of the \x2 minimum, with a bino-dominated LSP $m_{\widetilde{\chi}_1^0} = 181$~GeV, light stop $m_{\widetilde{t}_1} = 937$~GeV, gluino
$m_{\widetilde{g}} = 1133$~GeV, and heavy squark $m_{\widetilde{u}_L} = 1739$ GeV.

\begin{table}[ht]
  \small
    \centering
    \caption{Spectrum (in GeV) for $M_{1/2} = 708$~ GeV, $M_{V} = 2612$~GeV, $m_{t} = 174.4$~GeV, $\tan \beta$ = 21.83.
Here, $\Omega_{\rm CDM} h^2$ = 0.1110 and the lightest neutralino is greater than 99\% Bino.}
		\begin{tabular}{|c|c||c|c||c|c||c|c||c|c||c|c|} \hline
    $\widetilde{\chi}_{1}^{0}$&$145$&$\widetilde{\chi}_{1}^{\pm}$&$311$&$\widetilde{e}_{R}$&$264$&$\widetilde{t}_{1}$&$777$&$\widetilde{u}_{R}$&$1367$&$m_{h}$&$125.3$\\ \hline
    $\widetilde{\chi}_{2}^{0}$&$311$&$\widetilde{\chi}_{2}^{\pm}$&$1134$&$\widetilde{e}_{L}$&$750$&$\widetilde{t}_{2}$&$1259$&$\widetilde{u}_{L}$&$1489$&$m_{A,H}$&$1230$\\ \hline
    $\widetilde{\chi}_{3}^{0}$&$1131$&$\widetilde{\nu}_{e/\mu}$&$745$&$\widetilde{\tau}_{1}$&$153$&$\widetilde{b}_{1}$&$1224$&$\widetilde{d}_{R}$&$1420$&$m_{H^{\pm}}$&$1233$\\ \hline
    $\widetilde{\chi}_{4}^{0}$&$1133$&$\widetilde{\nu}_{\tau}$&$726$&$\widetilde{\tau}_{2}$&$733$&$\widetilde{b}_{2}$&$1348$&$\widetilde{d}_{L}$&$1491$&$\widetilde{g}$&$945$\\ \hline
		\end{tabular}
		\label{tab:masses}
\end{table}

\begin{table}[ht]
  \small
    \centering
    \caption{Spectrum (in GeV) for $M_{1/2} = 850$~ GeV, $M_{V} = 4310$~GeV, $m_{t} = 174.4$~GeV, $\tan \beta$ = 22.52.
Here, $\Omega_{\rm CDM} h^2$ = 0.1129 and the lightest neutralino is greater than 99\% Bino.}
		\begin{tabular}{|c|c||c|c||c|c||c|c||c|c||c|c|} \hline
    $\widetilde{\chi}_{1}^{0}$&$181$&$\widetilde{\chi}_{1}^{\pm}$&$382$&$\widetilde{e}_{R}$&$315$&$\widetilde{t}_{1}$&$937$&$\widetilde{u}_{R}$&$1596$&$m_{h}$&$124.7$\\ \hline
    $\widetilde{\chi}_{2}^{0}$&$382$&$\widetilde{\chi}_{2}^{\pm}$&$1301$&$\widetilde{e}_{L}$&$875$&$\widetilde{t}_{2}$&$1464$&$\widetilde{u}_{L}$&$1739$&$m_{A,H}$&$1413$\\ \hline
    $\widetilde{\chi}_{3}^{0}$&$1298$&$\widetilde{\nu}_{e/\mu}$&$872$&$\widetilde{\tau}_{1}$&$188$&$\widetilde{b}_{1}$&$1436$&$\widetilde{d}_{R}$&$1655$&$m_{H^{\pm}}$&$1416$\\ \hline
    $\widetilde{\chi}_{4}^{0}$&$1300$&$\widetilde{\nu}_{\tau}$&$848$&$\widetilde{\tau}_{2}$&$854$&$\widetilde{b}_{2}$&$1568$&$\widetilde{d}_{L}$&$1740$&$\widetilde{g}$&$1133$\\ \hline
		\end{tabular}
		\label{tab:massestwo}
\end{table}

The cumulative distribution function (CDF) percentage labeled on the right-hand axis of the \x2 plot in Figure~(\ref{fig:strip})
is a statistical tool that establishes the fraction of trials (for a fixed number of statistically independent variables) where
Gaussian fluctuation of each of variable will yield a net deviation from the null hypothesis that is not larger
than the corresponding $\chi^2$ value referenced on the left-hand axis.
The two-sided limit as adopted for the $\chi^2$ best-fit minimization against $M_{1/2}$ using
searches with visible excesses is appropriate when meaningful deviations may be anticipated in either direction away
from the median.  Particularly small $\chi^2$ values imply a better fit to the data by the assumed
signal (at a certain confidence level) than what could be attributed to random fluctuations around the SM, while excessively
large $\chi^2$ values disfavor a given $M_{1/2}$ relative to the SM-only null hypothesis.  In this case, the usual
$1$ and $2$-$\sigma$ $68\%$ and $95\%$ consistency integrations enclose areas symmetrically distributed about the mean,
such that centrally inclusive boundary lines are drawn at the CDF percentages $2.2\%$, $15.9\%$, $84.1\%$, and $97.7\%$.
It should be noted that this method implicitly assumes the displayed channels to be statistically uncorrelated, which
is not perfectly applicable in the current case.  A compensating reduction in the effective degrees of freedom
from the nominal value of three would have the effect of marginally lowering the quoted CDF scale values relative to
the left-hand $\chi^2$ axis, slightly compressing the displayed error margins.

\section{Multijets and Light Stops}

It is not, in our opinion, coincidental that the only search strategies to exhibit appreciable signal strength thus
far, amongst the still relatively small 5 \fb of accumulated statistics, consist of multijets. In fact, this
feature is handily in accordance with our expectations for
No-Scale \fsu5, as outlined in Refs.~\cite{Maxin:2011hy,Li:2011hr}. The model preference for multijet events results
from leveraging the flippon multiplets to facilitate the flat RGE running from $M_{32}$ to $M_V$, engendering the
distinctive mass texture $M({\widetilde{t_1}}) < M({\widetilde{g}}) < M({\widetilde{q}})$, featuring a light stop
and gluino that are both substantially lighter than all other squarks. The spectrum produces a characteristic event
topology starting with the pair production of heavy first or second generation squarks $\widetilde{q}$ and/or gluinos
$\widetilde{g}$ in the initial hard scattering process, with each heavy squark likely to yield a quark-gluino pair
$\widetilde{q} \rightarrow q \widetilde{g}$ in the cascade decay. Each gluino will most likely decay to light stops
$\widetilde{t}_1$ via $\widetilde{g} \rightarrow \widetilde{t}_1 \overline{t} \rightarrow t \overline{t} \widetilde{\chi}_1^0$,
or absent light stops, $\widetilde{g} \rightarrow q \overline{q} \widetilde{\chi}_1^0$, with each gluino producing 2--6 jets,
where the gluino-mediated stop channel produces the maximum of six jets.

In Refs.~\cite{Maxin:2011hy,Li:2011hr}, we advocated for a rather soft cut on hadronic jets of $p_T > 20$ GeV,
in order to capture as many large multijet events with at least nine jets as possible. While we are certainly pleased that
ATLAS has focused attention on multijet events, as evidenced by the powerful search strategies highlighted in
Figure~(\ref{fig:strip}), the $p_T$ cut on jets in these strategies is too hard to preserve events with at least nine
jets in only 5 \fb of 7 TeV statistics. Consequently, we find the very recent 5~\fb light stop search results in ATLAS Ref.~\cite{ATLAS-CONF-2012-073}
quite exciting, where a softer jet cut of $p_T > 25$ GeV was employed, producing a notable excess of pair-produced
light stop event candidates in the $E_T^{Miss}$ bin correlated to about twice the $m_{\chi} = 145$ GeV lightest neutralino
mass~\cite{Li:2011gh} of the best fit benchmark from Table~\ref{tab:masses}.  A close scrutiny of the ATLAS light stop
search reveals Figure (5) of Ref.~\cite{ATLAS-CONF-2012-073}, clearly illustrating that all of the 2.5$\sigma$ excess
events emanated from events with 9-10 jets in both the electron and muon channels, entirely consistent with our
SUSY discovery prescription of Refs.~\cite{Maxin:2011hy,Li:2011hr}.  We emphasize that the magnitude of the observed
excess~\cite{Li:2012hm,Li:2012tr} and the value of the missing energy at which any observed excess is maximized~\cite{Li:2011gh}
are independent discriminants in the \fsu5 context, which must each ultimately link fundamental SUSY processes back to
a single consistent gaugino mass $M_{1/2}$, the wellspring from which all correlations within this model flow.

A full-scale analysis of the ATLAS light stop search of Ref.~\cite{ATLAS-CONF-2012-073} in the context of an \fsu5
framework is in progress~\cite{LMNW-P}, and indeed we have already established that the best SUSY mass fit of the light
stop excesses are well correlated with the three multijet ATLAS searches depicted in Figure (\ref{fig:strip}). Moreover,
since the search strategy of Ref.~\cite{ATLAS-CONF-2012-073} targets pair-produced light stops, while the other three
searches of Figure (\ref{fig:strip}) target pair-produced gluinos and/or squarks, this light stop search should be
statistically independent of these other high-productivity searches. The SRC Tight (4 jets and $p_T > 60$ GeV for j2--j4),
SRE Loose (6 jets and $p_T > 60$ GeV for j2--j4, $p_T > 40$ GeV for j5--j6), and 7j80 ($\ge$7 jets and $p_T > 80$ GeV)
searches are also statistically independent as a result of the larger $p_T >$ 80 GeV cut for 7j80, such that to capture
the same event between more than one of these three searches would require lowering the number of jets in 7j80 events to
below seven, which would would in turn necessitate raising the $p_T$ cut to greater than 80 GeV, though the jet cut is lower
than 80 GeV for SRC Tight and SRE Loose.

\section{Fast Proton Decay\label{sct:protons}}

Instability of the proton is an essential signature of GUTs,
the merger of the Standard Model forces necessarily linking quarks to leptons,
and providing a narrow channel $p \!\rightarrow\! {(e\vert\mu)}^{\!+}\! \pi^0$
of dimension six decay via heavy gauge boson exchange.
The 50-kiloton (kt) water \v{C}erenkov detector of the Super-Kamiokande facility
has now set a lower bound of $1.4\times 10^{34}$ years for the partial lifetime in the
$p\rightarrow e^+ \pi^0$ mode, and somewhat less for $\mu^+ \pi^0$~\cite{Hewett:2012ns}.
The predicted lifetime for each of these two channels may be comparable in flipped $SU(5)$.
However, this construction evades the dangerously rapid $p\rightarrow K^+ {\bar \nu}$
dimension five triplet Higgsino mediated decay that plagues other SUSY GUTs 
by way of the missing-partner mechanism~\cite{Antoniadis:1987dx}, which
naturally facilitates Higgs doublet-triplet splitting without the side effect of strong triplet mixing.

We have previously~\cite{Li:2010dp} provided a comprehensive dictionary for the translation
in closed form of low energy experimental observables into their flipped $SU(5)$ counterparts
expressed at the high energy scale, accurate up to leading effects in the second order.
In particular, the two couplings $\alpha_{5}$ and $\alpha_{\rm X}$ of the flipped $SU(5)\times U(1)_{\rm X}$
gauge group and the GUT scale mass $M_{32}$ --- at which the partially unified $SU(5)$ sector decomposes
into $SU(3)\times SU(2)_{\rm L}$ and the Abelian phase associated with this breaking of rank from
4 to 3 likewise remixes with the quarantined $U(1)_{\rm X}$ subgroup to form the weak scale hypercharge
$U(1)_{\rm Y}$ --- may be written in terms of the Z-boson mass $M_{\rm Z}$, the values of the strong and
electromagnetic couplings ($\alpha_{\rm s}$, $\alpha_{\rm em}$) and Weinberg angle $\sin^2 \theta_{\rm W}$
at $M_{\rm Z}$, the top quark mass $m_{\rm t}$ and the low-energy SUSY particle spectrum (for computation
of threshold corrections), and the $\beta$-function renormalization group coefficients at the first and second
loops. Our current interest is a direct isolation of dependencies in $M_{32}$ and $1/\alpha_5$,
to which the proton lifetime $\tau_{\rm p}$
in the $p \!\rightarrow\! {(e\vert\mu)}^{\!+}\! \pi^0$ dimension six modes
is proportional in the fourth and second powers respectively,
upon variation of the strong coupling $\alpha_{\rm s}(M_{\rm Z})$.

\begin{figure}[htp]
\begin{center}
\includegraphics[width=0.5\textwidth,angle=0]{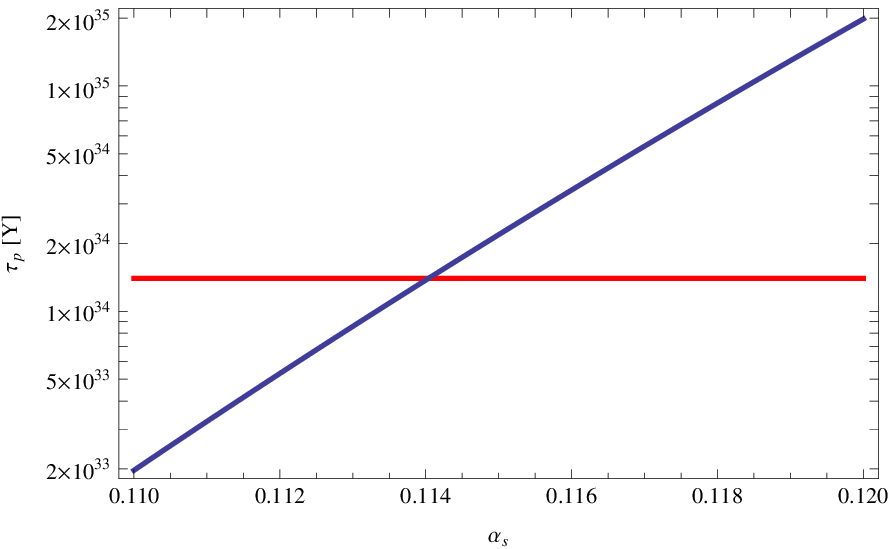}
\end{center}
\caption{\label{fig:tofas}
The variation of the proton lifetime with $\alpha_{\rm s}$ is depicted
by the blue curve. The red horizontal line establishes a lower bound
of $1.4\times10^{34}$~Y on the partial proton lifetime in the
$e^+ \pi^0$ channel.}
\end{figure}

To facilitate this effort, we borrow the following from Eq. Set~(69) of Ref.~\cite{Li:2010dp}.
\begin{eqnarray}
M_{32} &=&
M_{\rm Z} \times \exp \left\{
\frac{ 2\pi (\alpha_{\rm s} \Theta_{\rm W} - \alpha_{\rm em} \Xi_3)
}{
\alpha_{\rm em} \alpha_{\rm s} (b_2-b_3)}
\right\}
\nonumber \\
\frac{1}{\alpha_5} &=&
\frac{ \alpha_{\rm em} \Xi_3 b_2 - \alpha_{\rm s} \Theta_{\rm W} b_3
}{
\alpha_{\rm em} \alpha_{\rm s} (b_2-b_3) }
\label{eq:m32alpha5}
\end{eqnarray}
In the prior, \mbox{$\Theta_{\rm W} \equiv \sin^2\theta_{\rm W} + \alpha_{\rm em} \:(\xi_2-\zeta_2) / 2\pi$}
and \mbox{$\Xi_3 \equiv 1 + \alpha_{\rm s} \:(\xi_3-\zeta_3) / 2\pi$},
where $\xi_i$ and $\zeta_i$ respectively incorporate effects
associated to the $i^{\rm th}$ gauge coupling of mass threshold crossings
and the second quantum loop. The $b_i$ parameters are the single loop
$\beta$-function coefficients for the running of the $i^{\rm th}$ coupling.
The $\alpha_{i}$ to which we refer are the usual GUT-normalized
$SU(3)\times SU(2)_{\rm L} \times U(1)_{\rm Y}$ couplings, with boundaries expressed as follows.
\begin{eqnarray}
\alpha_1(M_{\rm Z}) &=&
\frac{5\, \alpha_{\rm em}}{3\,(1-\sin^2\theta_{\rm W})}
\nonumber \\
\alpha_2(M_{\rm Z}) &=& \frac{\alpha_{\rm em}}{\sin^2\theta_{\rm W}}
\nonumber \\
\alpha_3(M_{\rm Z}) &=& \alpha_{\rm s}
\end{eqnarray}
The crucial observation is that $\Theta_{\rm W}$ has no leading dependence on $\alpha_{\rm s}$,
while $\Xi_3$ represents a deviation from unity that is linearly proportional to $\alpha_{\rm s}$
at leading order. The residual dependency of Eq.~(\ref{eq:m32alpha5}) upon the strong coupling
is thus very well contained. It may be expressed as follows, where
$A$ and $B$ are factors that depend only weakly upon $\alpha_{\rm s}$,
and which may be set by normalization against a pair of known solutions.
\begin{eqnarray}
M_{32} &=& A \times \exp \left\{ \frac{ - 2\pi }{ \alpha_{\rm s} (b_2-b_3)} \right\}
\nonumber \\
\frac{1}{\alpha_5} &=& B + \frac{ b_2 }{ \alpha_{\rm s} (b_2-b_3) }
\label{eq:tofas}
\end{eqnarray}
The linearized variations may be read off immediately.
\begin{eqnarray}
\delta \left( M_{32} \right) &=& M_{32} \times \frac{ - 2\pi }{ (b_2-b_3)} \times \delta \left( \frac{1}{\alpha_{s}} \right)
\nonumber \\
\delta \left( \frac{1}{\alpha_5}\right) &=& \frac{ b_2 }{ (b_2-b_3) } \times \delta \left( \frac{1}{\alpha_{s}} \right)
\label{eq:linvars}
\end{eqnarray}
Intuitively, lifts in $1/\alpha_{\rm s}$ are transmitted rather
directly to $1/\alpha_5$, modulo some contribution from the differential
slopes of the $1/\alpha_2$ and $1/\alpha_3$ logarithmic running.
A larger value of $\alpha_{\rm s}$ will generally require a greater
mass scale difference to be spanned in the running, in order to close
the wider initial separation between $\alpha_2$ and $\alpha_3$; as expected,
this is manifest in an upward shift of the partial unification point $M_{32}$.
Using the quoted quartic and quadratic scaling of the proton lifetime $\tau_{\rm p}$ with each of
$M_{32}$ and $1/\alpha_5$, the linearized variation of this composite factor may
also be deduced.
\begin{equation}
\delta \left( \tau_{\rm p} \right) = \tau_{\rm p} \times \frac{2 \,b_2\, \alpha_5 - 8\pi}{(b_2 - b_3)} \times \delta \left( \frac{1}{\alpha_{s}} \right)
\label{eq:tvar}
\end{equation}
The specific $\beta$-function coefficients employed in this study are those corresponding to
the $\cal{F}$-$SU(5)$ model, as modified by the flippon vector-like multiplet field content.
\begin{equation}
b_2 = 4 \qquad;\qquad b_3 = 0
\label{eq:betas}
\end{equation}
The dominant effect in Eq.~(\ref{eq:tvar}) is thus the negative term associated with variation
of the $M_{32}$ mass scale.
In Figure~(\ref{fig:tofas}), we plot the variation of the proton lifetime with $\alpha_{\rm s}$
given by the approximation of Eq. Set~(\ref{eq:tofas}), {\it i.e.}~without feedback into the
mass thresholds or second loop tabulation.
The absolute numerical scaling of the proton lifetime is set relative to the following parameterization
which is specialized to flipped $SU(5)$~\cite{Murayama:2001ur,Ellis:1993ks,Li:2010dp}.
\begin{equation}
\tau_{p} = 3.8 \times
{\left( \frac{M_{32} }{10^{16}~{\textrm{[GeV]}}} \right)}^4
\times {\left( \frac{0.0412 }{ \alpha_5} \right)}^2 \times 10^{35}~{\textrm{[Y]}} \label{plife_flipped}
\end{equation}
The normalization used is relative to a complete calculation based on a phenomenologically favored
benchmark spectrum with a gaugino mass of $M_{1/2} = 708$~GeV, using $\alpha_{\rm s} = 0.1172$.
The results compare very favorably with a second complete calculation at the same point in the
model space, explicitly substituting $\alpha_{\rm s} = 0.1145$ at all intermediate steps,
corresponding to the spectrum provided in Table~(\ref{tab:masses}).
The isolated values of $M_{32}$ and $\alpha_5$ track to better than one percent throughout this
rather wide gap in $\alpha_{\rm s}$, and the proton lifetime agrees within about two percent.
The observed variation in the lifetime moreover agrees to a level of about 10 percent across
this span with the closed form linearized formula in Eq.~(\ref{eq:tvar}), given mean values
for the referenced $\tau_{\rm p}$ and $\alpha_{5}$ baselines. The tracking of $\delta(1/\alpha_5)$
from Eq. Set~(\ref{eq:linvars}) is less satisfactory, although its influence is subordinate to
that of $\delta(M_{32})$, for which the linearized formula is an excellent approximation.

We establish by this parameterization a lower bound of $\alpha_{\rm s} \ge 0.1140$ for the
strong coupling in the context of our favored benchmark $\cal{F}$-$SU(5)$ model, beneath
which the proton lifetime impinges upon the existing limit of $1.4\times 10^{34}$~years in the
$e^{\!+} \pi^0$ decay channel.  It should be emphasized that rather fast~\cite{Li:2010dp}
proton decay is a generic feature of the $\cal{F}$-$SU(5)$ model, due to intrinsic
strengthening of the relevant coupling ($\alpha_5$) and intrinsic lowering of the relevant
mass scale ($M_{32}$). The situation becomes all the more acute in the context of a
mildly depressed input for the strong coupling $\alpha_{\rm s}$, as favored
both by certain recent direct observations~\cite{Bandurin:2011sh} and by the predictions of the the No-Scale $\cal{F}$-$SU(5)$
model relative to the heaviness of the Higgs mass. The resulting proton lifetime
prediction descends in this case to near a point of contact with the current experimental reach,
taking a central value of $1.7\times10^{34}$~years for $\alpha_{\rm s} = 0.1145$.  Unaccounted
heavy threshold effects near the GUT scale will tend to lengthen rather than shorten this
lifetime, by a factor perhaps as large as four~\cite{Li:2010dp}.

\section{130 GeV Gamma-ray Line}

The observation of a 130 GeV gamma-ray line~\cite{Weniger:2012tx} emanating from our galactic center by
the FERMI-LAT Space Telescope has initiated investigations into whether such a monochromatic line could
be attributed to dark matter annihilations, an argument amplified by the lack of any known astrophysical
source capable of producing a tantamount signature. The lightest neutralino mass at the minimum of the \x2
fit to the ATLAS multijet and light stop excesses is $m_{\chi} = 145$ GeV, clearly highlighted as the
benchmark model in Figure (\ref{fig:strip}). Conjecturing the observed photon line originates from
neutralino annihilations into a $Z$-boson and gamma-ray via $\widetilde{\chi} \widetilde{\chi} \to Z
\gamma$, we can compute the kinematics for a non-relativistic lightest neutralino
$\widetilde{\chi}_1^0$ as
\begin{eqnarray}
E_{\gamma} = M_{\chi} - \frac{M_Z^2}{4 M_{\chi}} \, ,
\label{eq:E}
\end{eqnarray}
which gives
\begin{eqnarray}
M_{\chi} = \frac{E_{\gamma}}{2}\left[1 + \sqrt{1 + \left(\frac{M_Z}{E_{\gamma}}\right)^2}\right]
\label{eq:M}
\end{eqnarray}
Using $E_{\gamma} \simeq 130$ GeV and $M_Z = 91.187$ GeV, we arrive at
\begin{eqnarray}
M_{\chi} = 144.4~ {\rm GeV} \, ,
\nonumber
\label{eq:W}
\end{eqnarray}
which is consistent with the No-Scale \fsu5 lightest neutralino mass of $M_{\chi} = 145$ GeV
at the best fit to the multijet and light stop excesses at the LHC $and$ generates an $m_h \simeq 125.5$ GeV Higgs boson mass.

The \fsu5 lightest neutralino composition is greater than 99\% bino, therefore, we cannot neglect the
quite small $\widetilde{\chi} \widetilde{\chi} \to Z \gamma$ annihilation cross-section
$\left\langle \sigma v \right\rangle \sim 10^{-30}~{\rm cm^3/s}$, about three orders of magnitude less
than the FERMI-LAT telescope observations. On the other hand, a recent analysis~\cite{Hektor:2012kc}
of extra-galactic clusters uncovering synonymous 130 GeV gamma-ray lines has determined an appropriate
subhalo boost factor in this context of $\sim 1000$ relative to the galactic center.  We do not consider it
implausible that an overall unaccounted boost factor of similar magnitude might reconcile this apparent discrepancy
of scale.  For now, we are content to simply make note of the interesting correlation that exists between 
145 GeV \fsu5 neutralino annihilations and the unexplained 130 GeV gamma-ray line observed by
the FERMI-LAT space telescope, irrespective of the absolute $\left\langle \sigma v \right\rangle$ 
cross-section magnitude. 

\section{Imminent Testability}

The full set of correlations enumerated here are imminently testable. We itemized in
Ref.~\cite{Li:2012qv} the projected signal significance of the ATLAS multijet SUSY searches for the
already accumulated 8 TeV collision data, suggesting that indeed an \fsu5 framework presently under
probe by the LHC should show evidence in the first 6 \fb of 8 TeV data amassed in 2012 of signal significances
approaching and possibly breaching the gold standard of \sb $\ge$ 5. This tantalizing production of multijet events beyond the
Standard Model background has increased proportionally from 1 \fb to 5 \fb in the 7 TeV collision data,
diminishing the prospect of random background fluctuations as the origin of the excess
events~\cite{Li:2012qv}.

We have detailed here the interconnectivity implied by the recently measured $m_h \simeq 125.5 \pm
0.5$ GeV Higgs boson mass and the ATLAS SUSY searches at the LHC.  We now append
to the markers of imminent testability for No-Scale \fsu5
i) a readily observable proton lifetime, perhaps as low as $1.7\times10^{34}$~years,
which is certainly within the reach of the next
generation Hyper-Kamiokande and DUSEL experiments~\cite{Nakamura:2003hk,Kearns:2008,Raby:2008pd}
and already near the grasp of Super-Kamiokande~\cite{Hewett:2012ns},
and ii) a 145 GeV WIMP mass potentially associable with the 130 GeV gamma-ray line observations by
the FERMI-LAT Space Telescope.

Whereas the landscape of supersymmetric models is replete with predictions requiring years more of massive data
observations and significantly higher LHC beam energies, we emphasize here to the contrary that
all the No-Scale \fsu5 predictions detailed in this work dwell just on the cusp of an
experimentally significant discovery.
The mutual correspondence exhibited across multiple independent channels of empirical evidence with
the No-Scale \fsu5 high-energy framework heightens the potentiality that the coming year could
present a complementary discovery to the Higgs boson of equal or even more profound consequence.


\begin{acknowledgments}
We acknowledge the generous contribution of Chunli Tong and Yunjie Huo of the
State Key Laboratory of Theoretical Physics in Beijing for
their calculation of Eqs.~(\ref{Delta mhs}-\ref{eq:higgs}). This research was supported in part
by the DOE grant DE-FG03-95-Er-40917 (TL and DVN),
by the Natural Science Foundation of China
under grant numbers 10821504, 11075194, and 11135003 (TL),
and by the Mitchell-Heep Chair in High Energy Physics (JAM).
We also thank Sam Houston State University
for providing high performance computing resources.
\end{acknowledgments}


\bibliography{bibliography}

\end{document}